\documentclass[preprintnumbers,amsmath,amssymb,floatfix,12pt,prd,superscriptaddress,nofootinbib]{revtex4}
\usepackage{graphicx}
\usepackage{epsfig}
\usepackage{bm}
\usepackage{amsfonts}

\begin{document}

\title{Irreversible thermodynamics of dark energy on the
entropy-corrected apparent horizon}

\author{K. Karami}
\email{KKarami@uok.ac.ir} \affiliation{Department of Physics,
University of Kurdistan, Pasdaran St., Sanandaj, Iran}
\affiliation{Research Institute for Astronomy $\&$ Astrophysics of
Maragha (RIAAM), Maragha, Iran}

\author{M. Jamil}
\email{mjamil@camp.nust.edu.pk} \affiliation{Center for Advanced
Mathematics and Physics (CAMP), National University of Sciences and
Technology (NUST), Islamabad, Pakistan}

\author{N. Sahraei}
\affiliation{Department of Physics, University of Kurdistan,
Pasdaran St., Sanandaj, Iran}

\date{\today}

\begin{abstract}
\vspace*{1.5cm} \centerline{\bf Abstract} \vspace*{0.5cm} We study
the irreversible (non-equilibrium) thermodynamics of FRW universe
containing only dark energy. Using the modified entropy-area
relation which is motivated from the loop quantum gravity, we
calculate the entropy-corrected form of the apparent horizon of FRW
universe.
\end{abstract}

\maketitle


\clearpage
\section{Introduction}
Observations of type Ia supernovae suggest that the universe is
dominated by two dark components: dark matter (DM) and dark energy
(DE) \cite{Riess}. Dark matter, a matter without pressure, is mainly
used to explain galactic rotation curves and the formation of
large-scale structure, while dark energy (DE), an exotic energy with
negative pressure, is used to describe the present cosmic
accelerated expansion. However, the nature and origin of DE is still
unknown, and people have proposed several candidates to describe it
(for review see \cite{Karami1} and references therein).

The holographic DE (HDE) is one of interesting DE candidates which
was proposed based on the holographic principle \cite{Horava}.
According to the holographic principle, the number of degrees of
freedom in a bounded system should be finite and has relationship
with the area of the boundary of the system \cite{Hooft}. By
applying the holographic principle to cosmology, one can obtain
the upper bound on the entropy contained in the universe
\cite{Fischler}. Following this line, Li \cite{Li} argued that in
quantum field theory, the ultra-violet cut-off $\Lambda $ could be
related to the infrared cut-off $L$ due to the limit set by
forming a black hole i.e. the quantum zero-point energy of a
system with size $L$ should not exceed the mass of a black hole
with the same size, i.e. $L^{3}\Lambda
^{3}\leq (M_{p}L)^{3/2}.$ This last expression can be re-written as $%
L^{3}\rho _{\Lambda }\leq LM_{p}^{2},$ where $\rho _{\Lambda
}\backsim \Lambda ^{4}$ is the energy density corresponding to the
zero point energy and cut-off $\Lambda .$ Now the last inequality
takes the form $\rho _{\Lambda }\leq M_{p}^{2}L^{-2}$ or $\rho
_{\Lambda }=3c^{2}M_{p}^{2}L^{-2}.$ Here $M_{p}^{2}=(8\pi G)^{-1}$
is the modified Planck mass and $3c^{2}$ is constant and attached
for convenience. The HDE models have been studied widely in the
literature \cite{Enqvist,Elizalde2,Guberina}. Obviously, in the
derivation of HDE, the black hole entropy $S_{\rm BH}$ plays an
important role. As is well known $S_{\rm BH} = A/(4G)$, where
$A\sim L^2$ is the area of horizon. However, in the literature,
this entropy-area relation can be modified to \cite{modak}
\begin{equation}
S_{\rm
BH}=\frac{A}{4G}+\tilde{\alpha}\ln{\frac{A}{4G}}+\tilde{\beta},\label{MEAR}
\end{equation}
where $\tilde{\alpha}$ and $\tilde{\beta}$ are dimensionless
constants of order unity. These corrections can appear in the
black hole entropy in loop quantum gravity (LQG) \cite{HW}. They
can also be due to thermal equilibrium fluctuation, quantum
fluctuation, or mass and charge fluctuations (for a good review
see \cite{HW} and references therein).

An interesting aspect in the cosmological context is the study of
non-equilibrium thermodynamics of DE. Das et al. \cite{Das}
computed leading-order corrections to the entropy of any
thermodynamic system due to small statistical fluctuations around
equilibrium. They obtained a general logarithmic correction to
black hole entropy. Wang et al. \cite{Wang4} studied a
thermodynamical description of the interaction between HDE and DM.
Resorting to the logarithmic correction to the equilibrium
entropy, they
 arrived to an expression for the interaction term which was
consistent with the observational tests. Pav\'{o}n and Wang
\cite{Pavon} considered a system composed of two subsystems (DM and
DE) at different temperatures. In virtue of the extensive property,
the entropy of the whole system is the sum of the entropies of the
individual subsystems which (being equilibrium entropies) are just
functions of the energies of DE and DM. Zhou et al. \cite{Zhou1}
have further employed the second law of thermodynamics to study the
coupling between the DE and DM in the universe by resorting to the
non-equilibrium entropy of extended irreversible thermodynamics.
Karami and Ghaffari \cite{Karami2} extended the work of Zhou et al.
\cite{Zhou1} and investigated the validity of the generalized second
law in irreversible thermodynamics for the interacting DE with DM in
a non-flat FRW universe enclosed by the dynamical apparent horizon.
It was shown that for the present time, the generalized second law
in non-equilibrium thermodynamics is satisfied for the special range
of the energy transfer constants. More recently, Gang and Wen-Biao
\cite{WG} studied the non-equilibrium thermodynamics of DE on cosmic
apparent horizon. They clarified that if the irreversible process is
considered, the proper position for building thermodynamics will not
be the apparent horizon anymore. The new position is related to DE
state equation and the irreversible process parameters.

In this paper, our aim is to extend the work of Gang and Wen-Biao
\cite{WG} using the quantum corrected entropy-area relation
\cite{HW}. We investigate how the original apparent horizon can be
modified to keep the non-equilibrium thermodynamic laws in effect.
This paper is organized as follows. In Section 2, we study the
non-equilibrium thermodynamics of DE on entropy-corrected apparent
horizon. Section 3 is devoted to the conclusions.
\section{Non-equilibrium thermodynamics of DE on entropy-corrected apparent
horizon}

We consider the Friedmann-Robertson-Walker (FRW) metric for the
non-flat universe as
\begin{equation}
{\rm d}s^2=-{\rm d}t^2+a^2(t)\left(\frac{{\rm
d}r^2}{1-kr^2}+r^2{\rm d}\Omega^2\right),\label{metric}
\end{equation}
where $k=0,1,-1$ represents a spatially flat, closed or open FRW
universe, respectively. Define $\tilde{r}(t) = a(t)r$, the metric
(\ref{metric}) can be rewritten as ${\rm d}s^2 = h_{ij}{\rm
d}x^i{\rm d}x^j + \tilde{r}^2{\rm d}\Omega^2$, where $x^i = (t, r)$,
$h_{ij}$ = diag($-1, a^2/(1 - kr^2)$), $i,j=0,1$. By definition
\begin{equation}
f:=h^{ij}\partial_{i}\tilde{r}\partial_{j}\tilde{r}=1-\left(H^2+\frac{k}{a^2}\right)\tilde{r}^2
,\label{f}
\end{equation}
when $f=0$ then the location of the apparent horizon in the FRW
universe is obtained as $\tilde{r} = R_{\rm A} =
(H^2+k/a^2)^{-1/2}$ \cite{Poisson}. Observational evidences
suggest that our universe is spatially flat with $k = 0$
\cite{Bennett}, so the apparent horizon is same as the Hubble
horizon, i.e. $R_{\rm A} = 1/H$.

With the help of Eq. (\ref{f}), the surface gravity $\kappa$ and
the Hawking temperature $T$ are defined as \cite{Gong}
\begin{equation}
\kappa=-\frac{\partial_{\tilde{r}}f}{2}=\frac{\tilde{r}}{R_{\rm
A}^2},
\end{equation}
\begin{equation}
T= \frac{\kappa}{2\pi}=\frac{\tilde{r}}{2\pi R_{\rm A}^2}.
\end{equation}
So the Hawking temperature at the apparent horizon is obtained as
\begin{equation}
T_{\rm A} = T\Big|_{\tilde{r}=R_{\rm A}}=\frac{1}{2\pi R_{\rm A}}.
\end{equation}
Recently Cai et al. \cite{Cai09} proved that the apparent horizon
of the FRW universe has an associated Hawking temperature $T_{\rm
A} = 1/2\pi R_{\rm A}$. They also showed that this temperature can
be measured by an observer with the Kodoma vector inside the
apparent horizon.

The first Friedmann equation for the flat FRW universe takes the
form
\begin{equation}
H^2=\frac{8\pi}{3}\rho,\label{eqfr}
\end{equation}
where we take $G=1$ and $\rho$ is the energy density of DE. The
continuity equation for DE is
\begin{equation}
\dot{\rho}+3H(1+\omega)\rho=0,\label{eq}
\end{equation}
where $\omega$ is the parameter of the equation of state (EoS) of
DE, and when $\omega<-1/3$ it describes an accelerating universe.
For a constant $\omega$, from Eqs. (\ref{eqfr}) and (\ref{eq}) we
get
\begin{equation}
a(t)=t^{1/\epsilon},~~~\epsilon:=\frac{3}{2}(1+\omega),\label{a}
\end{equation}
and the apparent (Hubble) horizon is obtained as $R_{\rm
A}=1/H=\epsilon t$. Note that to have $R_{\rm A}$ be positive, it
is also needed to assume $\epsilon>0$ or $\omega>-1$. Therefore
the DE with $-1<\omega<-1/3$ behaves like a quintessence scalar
field model \cite{Wetterich}. The flux of energy across the
apparent horizon within the time interval ${\rm d}t$ is
\begin{equation}
-{\rm d}E_{\rm A}=4\pi R_{\rm A}^2\rho(1+\omega){\rm
d}t=\epsilon{\rm d}t.
\end{equation}
The process of energy flux crossing apparent horizon is
irreversible. Therefore in the presence of interaction between DE
and its surroundings enveloped by the horizon, the time derivative
of the non-equilibrium entropy is given by
\begin{equation}
\dot{S}=\dot{S}_{\rm i}+\dot{S}_{\rm e},
\end{equation}
where $\dot{S}_{\rm i}$ is the rate of change in internal entropy
production of the universe and $\dot{S}_{\rm e}$ arises from the
heat flow between the universe and the horizon. Following \cite{WG}
we have
\begin{equation}
\dot{S}_{\rm i}=\int_V{{\sigma}{\rm d} V},\label{eqin1}
\end{equation}
\begin{equation}
\dot{S}_{\rm e}=-\oint_{\Sigma}\vec{ J_{S}}\cdot{\rm
d}\vec{\Sigma},\label{eqex1}
\end{equation}
where $\sigma$ is an internal entropy source production density and
$J_S$ is an entropy flow density. If we consider only the heat
conduction between the universe and the horizon, then $\sigma$ and
$J_S$ are given by \cite{WG}
\begin{equation}
\sigma=\vec{J_q}\cdot\nabla\frac{1}{T}\Big|_{\tilde{r}=R_{\rm
A}},\label{eqent}
\end{equation}
\begin{equation}
\vec{J_{S}}=\frac{\vec{J_q}}{T_{\rm A}}.\label{eqefd}
\end{equation}
Here $\vec{J_q}$ is the heat current. Substitution of Eq.
(\ref{eqefd}) in (\ref{eqex1}) yields
\begin{equation}
\dot{S}_{\rm e}=-\frac{1}{T_{\rm A}}\oint_{\Sigma}\vec{
J_q}\cdot{\rm d}\vec{\Sigma}=\frac{1}{T_{\rm A}}{J_q}{4\pi}{
R_{\rm A}^2},\label{eqex2}
\end{equation}
where we assume $\vec{J_q}$ takes the same value at every point of
the apparent horizon surface $A=4\pi{R_{\rm A}^2}$.

From the modified entropy-area relation (\ref{MEAR}) one obtains
\begin{equation}
\frac{\rm d}{{\rm d}t}\left(\frac{A}{4}+\tilde{\alpha}\ln\frac{
A}{4}+\tilde{\beta}\right)={2\pi}R_{\rm A}\dot{R_{\rm
A}}+\frac{2\tilde{\alpha}}{R_{\rm A}}\dot{R_{\rm A}}={2\pi} R_{\rm
A}\epsilon\left(1+\frac{\tilde{\alpha}}{{\pi} R_{\rm
A}^2}\right),\label{eqex3}
\end{equation}
where $\dot{R_{\rm A}}=\epsilon$. Like \cite{WG} thinking of the
relation
\begin{equation}
\dot{S}_{\rm e}=\frac{\rm d}{{\rm d}t}\left(\frac{
A}{4}+\tilde{\alpha}\ln\frac{A}{4}+\tilde{\beta}\right),\label{Sdote}
\end{equation}
from Eqs. (\ref{eqex2}) and (\ref{eqex3}), one gets
\begin{equation}
J_q=\frac{\epsilon}{{4\pi}{ R_{\rm
A}^2}}\left(1+\frac{\tilde{\alpha}}{{\pi} R_{\rm
A}^2}\right).\label{eqhc1}
\end{equation}
Following the Fourier's law
\begin{equation}
\vec{J_q}=-\lambda\nabla{T}\Big|_{\tilde{r}=R_{\rm
A}},\label{eqhc2}
\end{equation}
where $\lambda$ is the thermal conductivity \cite{WG}. This shows
that there will be spontaneous heat flow between the horizon and
the DE and the thermal equilibrium will no longer hold
\cite{Karami2}.

Substituting Eqs. (\ref{eqhc1}) and (\ref{eqhc2}) in (\ref{eqent})
gives
\begin{equation}
\sigma=\frac{\epsilon^2}{{4\lambda} R_{\rm
A}^2}{\left(1+\frac{\tilde{\alpha}}{{\pi}R_{\rm A}^2}\right)}^2,
\end{equation}
while from Eq. (\ref{eqin1}) we obtain
\begin{equation}
\dot{S}_{\rm i}=\sigma{\frac{4}{3}\pi{ R_{\rm
A}^3}}=\frac{\epsilon^2\pi{ R_{\rm
A}}}{3\lambda}{\left(1+\frac{\tilde{\alpha}}{{\pi }R_{\rm
A}^2}\right)}^2.\label{Sdoti}
\end{equation}
Finally from Eqs. (\ref{Sdote}) and (\ref{Sdoti}), the time
derivative of the irreversible entropy can be obtained as
\begin{equation}
\dot{S}=\dot{S}_{\rm i}+\dot{S}_{\rm e}=\frac{\epsilon^2\pi{ R_{\rm
A}}}{3\lambda}{\left(1+\frac{\tilde{\alpha}}{{\pi} R_{\rm
A}^2}\right)}^2+{2\pi}R_{\rm
A}\epsilon\left(1+\frac{\tilde{\alpha}}{{\pi} R_{\rm A}^2}\right).
\end{equation}
Therefore the first law in irreversible thermodynamics holds if we
define
\begin{equation}
{\rm d}S=-\frac{{\rm d}E_{\rm A}}{\tilde{T}_{\rm
A}}=2\pi\tilde{R}_{\rm A}\epsilon{\rm d}t,
\end{equation}
where $\tilde{T}_{A}:=1/2\pi\tilde{R}_{\rm A}$ and
\begin{equation}
\tilde{R}_{\rm A}={R_{\rm A}}\left(1+\frac{\tilde{\alpha}}{{\pi}
R_{\rm
A}^2}\right)\left[1+\left(\frac{\epsilon}{6\lambda}\right)\left(1+\frac{\tilde{\alpha}}{{\pi}
R_{\rm A}^2}\right)\right],
\end{equation}
which is so-called the entropy-corrected apparent horizon. If we set
$\tilde{\alpha}=0$ then
\begin{equation}
\tilde{R}_{\rm A}={R_{\rm
A}}\left(1+\frac{\epsilon}{6\lambda}\right),
\end{equation}
which is same as the result obtained by \cite{WG}.
\section{Conclusions}

In this paper, we investigated the FRW universe as a
non-equilibrium (or irreversible) thermodynamical system by
considering the logarithmic correction term to the horizon
entropy. We assumed that the universe contains only DE in dominant
form compared to other cosmic components, moreover, the universe
is bounded by apparent horizon. In a non-equilibrium scenario,
there is a flux of energy across the horizon. As energy goes
outside the horizon (and cannot come in due to irreversibility),
an internal entropy production term is taken into account. Notice
that this term is zero in the equilibrium setting. Recently Gang
and Wen-Biao \cite{WG} studied this problem and showed that when
the irreversible process is considered, the original apparent
horizon is no longer perfect for building non-equilibrium
thermodynamics laws exactly and it should be modified. They
obtained the modified apparent horizon expression that depended on
the state parameter of DE $\epsilon$ and a non-equilibrium factor
$\lambda$. We have extended their study by considering the
`corrected' entropy motivated from the LQG. The addition of
correction terms to the horizon entropy is a fundamental
prediction of LQG and hence must be taken into account while
studying the thermodynamics of horizons for the background
geometry. Using this modified entropy-area relation, we obtained
the entropy-corrected form of the apparent horizon of FRW
universe.
\begin{acknowledgments}
The authors thank the anonymous referee for very valuable
comments. The work of K. Karami has been supported financially by
Research Institute for Astronomy $\&$ Astrophysics of Maragha
(RIAAM), Maragha, Iran.
\end{acknowledgments}


\begin{thebibliography}{99}

\bibitem{Riess} A.G. Riess, et al., Astron. J. {\bf 116}, 1009 (1998);\\ S. Perlmutter, et
al., Astrophys. J. {\bf 517}, 565 (1999);\\ P. de Bernardis, et
al., Nature {\bf 404}, 955 (2000);\\ S. Perlmutter, et al.,
Astrophys. J. {\bf 598}, 102 (2003).

\bibitem{Karami1} T. Padmanabhan, Phys. Rep. {\bf 380}, 235 (2003);\\
P.J.E. Peebles, B. Ratra, Rev. Mod. Phys. {\bf 75}, 559 (2003);\\
E.J. Copeland, M. Sami, S. Tsujikawa, Int. J. Mod. Phys. D {\bf
15}, 1753 (2006);\\K. Karami, S. Ghaffari, J. Fehri, Eur. Phys. J.
C {\bf 64}, 85 (2009).

\bibitem{Horava} P. Horava, D. Minic, Phys. Rev. Lett. {\bf 85}, 1610 (2000);\\
 P. Horava, D. Minic, Phys. Rev. Lett. {\bf 509}, 138 (2001);\\
S. Thomas, Phys. Rev. Lett. {\bf 89}, 081301 (2002).

\bibitem{Hooft} G. 't Hooft, gr-qc/9310026;\\ L. Susskind, J. Math. Phys. {\bf
36}, 6377 (1995).

\bibitem{Fischler} W. Fischler, L. Susskind, hep-th/9806039.

\bibitem{Li} M. Li, Phys. Lett. B {\bf 603}, 1 (2004).

\bibitem{Enqvist} M. Jamil, et al., arXiv:1006.5365; \\
M.R. Setare, M. Jamil, Phys. Lett. B {\bf 690}, 1 (2010);\\
M.R. Setare, M. Jamil, J. Cosmol. Astropart.
Phys. {\bf 02}, 010 (2010);\\
M. Jamil, E.N. Saridakis, M.R. Setare, Phys. Lett. B {\bf 679},
172 (2009);\\ K. Enqvist, M.S. Sloth, Phys. Rev. Lett. {\bf 93},
221302 (2004);\\Q.G. Huang, Y. Gong, J. Cosmol. Astropart. Phys.
{\bf 08}, 006 (2004);\\ Q.G. Huang, M. Li, J. Cosmol. Astropart.
Phys. {\bf 08}, 013 (2004);\\ Y. Gong, Phys. Rev. D {\bf 70},
064029 (2004).

\bibitem{Elizalde2} E. Elizalde, S. Nojiri, S.D. Odintsov, P. Wang, Phys. Rev. D {\bf
71}, 103504 (2005);\\ B. Guberina, R. Horvat, H. Stefancic, J.
Cosmol. Astropart. Phys. {\bf 05}, 001 (2005);\\ B. Wang, E.
Abdalla, R.K. Su, Phys. Lett. B {\bf 611}, 21 (2005);\\ J.Y. Shen,
B. Wang, E. Abdalla, R.K. Su, Phys. Lett. B {\bf 609}, 200 (2005);\\
X. Zhang, F.Q. Wu, Phys. Rev. D {\bf 72}, 043524 (2005).

\bibitem{Guberina} A. Sheykhi, Class. Quant. Grav. {\bf 27},  025007 (2010);\\
A. Sheykhi, Phys. Lett. B {\bf 682}, 329 (2010);\\
 B. Guberina, R. Horvat, H. Nikolic, Phys. Lett. B {\bf 636}, 80
(2006);\\ H. Li, Z.K. Guo, Y.Z. Zhang, Int. J. Mod. Phys. D {\bf
15}, 869 (2006);\\ J.P.B. Almeida, J.G. Pereira, Phys. Lett. B
{\bf 636}, 75 (2006);\\   X. Zhang, Phys. Rev. D {\bf 74}, 103505
(2006);\\ X. Zhang, F.Q. Wu, Phys. Rev. D {\bf 76}, 023502
(2007);\\ L. Xu, J. Cosmol. Astropart. Phys. {\bf 09}, 016
(2009);\\ K. Karami, J. Fehri, Phys. Lett. B {\bf 684}, 61
(2010);\\ K. Karami, J. Cosmol. Astropart. Phys. {\bf 01}, 015
(2010);\\ K. Karami, A. Abdolmaleki, Phys. Scr. {\bf 81}, 055901
(2010);\\K. Karami, arXiv:1002.0431.


\bibitem{modak} H.M. Sadjadi, M. Jamil, arXiv:1002.3588;\\
S.W. Wei, Y.X. Liu, Y.Q. Wang, H. Guo,
arXiv:1002.1550;\\
 D.A. Easson, P.H. Frampton, G.F. Smoot, arXiv:1003.1528;\\
 M. Jamil, M.U. Farooq, J. Cosmol. Astropart.
Phys. {\bf 03}, 001 (2010);\\
 R. Banerjee, B.R. Majhi, Phys. Lett. B {\bf 662}, 62 (2008);\\
 R. Banerjee, B.R. Majhi, J. High Energy Phys. {\bf 06}, 095 (2008);\\ B.R.
Majhi, Phys. Rev. D {\bf 79}, 044005 (2009);\\ R. Banerjee, S.K.
Modak, J. High Energy Phys. {\bf 05}, 063 (2009);\\ S.K. Modak,
Phys. Lett. B {\bf 671}, 167 (2009).

\bibitem{HW} H. Wei, Commun. Theor. Phys. {\bf 52}, 743 (2009).

\bibitem{Das} S. Das, P. Majumdar, R.K. Bhaduri, Class. Quantum Grav.
{\bf 19}, 2355 (2002).

\bibitem{Wang4} B. Wang, C.Y. Lin, D. Pav\'{o}n, E. Abdalla, Phys.
Lett. B {\bf 662}, 1 (2008).

\bibitem{Pavon} D. Pav\'{o}n, B. Wang, Gen. Relativ. Gravit. {\bf 41}, 1
(2009).

\bibitem{Zhou1} J. Zhou, B. Wang, D. Pav\'{o}n, E. Abdalla, Mod.
Phys. Lett. A {\bf 24}, 1689 (2009).

\bibitem{Karami2} K. Karami, S. Ghaffari, Phys. Lett. B {\bf 685}, 115 (2010).

\bibitem{WG} W. Gang, L. Wen-Biao, Commun. Theor. Phys. {\bf 52}, 383 (2009).

\bibitem{Poisson} E. Poisson, W. Israel, Phys. Rev. D {\bf 41}, 1796 (1990);\\
S.A. Hayward, Phys. Rev. D {\bf 53}, 1938 (1996);\\ Y.G. Gong, A.
Wang, Phys. Rev. Lett. {\bf 99}, 211301 (2007).

\bibitem{Bennett} C.L. Bennett, et al., Astrophys. J. Suppl. {\bf 148}, 1 (2003);\\ D.N.
Spergel, Astrophys. J. Suppl. {\bf 148}, 175 (2003);\\ M. Tegmark,
et al., Phys. Rev. D {\bf 69}, 103501 (2004);\\ U. Seljak, A.
Slosar,
P. McDonald, J. Cosmol. Astropart. Phys. {\bf 10}, 014 (2006);\\
D.N. Spergel, et al., Astrophys. J. Suppl. {\bf 170}, 377 (2007).

\bibitem{Gong} Y. Gong, B. Wang, A. Wang, J. Cosmol. Astropart.
Phys. {\bf 01}, 024 (2007).

\bibitem{Cai09} R.G. Cai, L.M. Cao, Y.P. Hu, Class. Quantum Grav. {\bf 26}, 155018
(2009).

\bibitem{Wetterich} C. Wetterich, Nucl. Phys. B {\bf 302}, 668
(1988);\\ B. Ratra, J. Peebles, Phys. Rev. D \textbf{37},
3406–3427 (1988).

\end{thebibliography}
\end{document}